\documentclass[trackchanges]{aastex7}
\DeclareUnicodeCharacter{02BC}{\textquotesingle}

\begin{document}

\title{Detecting Galactic Rings in the DESI Legacy Imaging Surveys with Semi-Supervised Deep Learning}

\author[gname=Jianzhen, sname=Chen]{Jianzhen Chen} 
\affiliation{Shanghai Key Lab for Astrophysics, Shanghai Normal University, Shanghai 200234, People’s Republic of China}
\email{jzchen@shnu.edu.cn}

\author[orcid=0009-0009-1617-8747,sname=Luo]{Zhijian Luo}
\affiliation{Shanghai Key Lab for Astrophysics, Shanghai Normal University, Shanghai 200234, People’s Republic of China}
\email[show]{zjluo@shnu.edu.cn}  

\author[orcid=0000-0003-0202-0534,gname=Cheng,sname=Cheng]{Cheng Cheng}
\affiliation{Chinese Academy of Sciences South America Center for Astronomy, National Astronomical Observatories, CAS, Beijing, 100101, Peopleʼs Republic of China}
\affiliation{National Astronomical Observatories, Chinese Academy of Sciences, 20A Datun Road, Chaoyang District, Beijing 100101, Peopleʼs Republic Of China}
\affiliation{CAS Key Laboratory of Optical Astronomy, National Astronomical Observatories, Chinese Academy of Sciences, Beijing 100101, Peopleʼs Republic of China}
\email{chengcheng@nao.ac.cn}

\author{Jun Hou}
\affiliation{Shanghai Key Lab for Astrophysics, Shanghai Normal University, Shanghai 200234, People’s Republic of China}
\email{junhou@shnu.edu.cn}

\author[0000-0002-2326-0476,sname=Zhang,gname=Shaohua]{Shaohua Zhang}
\affiliation{Shanghai Key Lab for Astrophysics, Shanghai Normal University, Shanghai 200234, People’s Republic of China}
\email{zhangshaohua@shnu.edu.cn}

\author[gname=Chenggang]{Chenggang Shu}
\affiliation{Shanghai Key Lab for Astrophysics, Shanghai Normal University, Shanghai 200234, People’s Republic of China}
\email{cgshu@shao.ac.cn}

\begin{abstract}

The ring structures of disk galaxies are vital for understanding galaxy evolution and dynamics. However, due to the scarcity of ringed galaxies and challenges in their identification, traditional methods often struggle to efficiently obtain statistically significant samples. To address this, this study employs a novel semi-supervised deep learning model, GC-SWGAN, aimed at identifying galaxy rings from high-resolution images of the DESI Legacy Imaging Surveys. We selected over 5,000 confirmed ringed galaxies from the Catalog of Southern Ringed Galaxies (CSRG) and the Northern Ringed Galaxies from the GZ2 catalog (GZ2-CNRG), both verified by morphology expert R. J. Buta, to create an annotated training set. Additionally, we incorporated strictly selected non-ringed galaxy samples from the Galaxy Zoo 2 dataset and utilized unlabelled data from DESI Legacy
Surveys to train our model. Through semi-supervised learning, the model significantly reduced reliance on extensive annotated data while enhancing robustness and generalization. On the test set, it demonstrated exceptional performance in identifying ringed galaxies. With a probability threshold of 0.5, the classification accuracy reached 97\%, with precision and recall for ringed galaxies at 94\% and 93\%, respectively. Building on these results, we predicted 750,000 galaxy images from the DESI Legacy Imaging Surveys with r-band apparent magnitudes less than 17.0 and redshifts in the range $0.0005 < z < 0.25$, compiling the largest catalog of ringed galaxies to date, containing 62,962 galaxies with ring structures. This catalog provides essential data for subsequent research on the formation mechanisms and evolutionary history of galaxy rings.

\end{abstract}



\section{Introduction} \label{sec:intro}

A crucial aspect of understanding galaxy evolution is the systematic study of their morphological features. These features, such as spiral arms, bars, ring structures, and tidal tails, not only serve as important clues for investigating galaxy structure, origin, and evolution but also reflect the dynamical states of galaxies and reveal their interactions and evolutionary processes within the cosmic environment \citep{lambas2003galaxy,lambas2012galaxy,alonso2012galaxy,vollmer2013large,almeida2014metallicity,mesa2014interacting,ho2019gas}. Therefore, systematic research on these morphological features is of great significance.

Galaxy rings, as a prominent structural feature, have been widely observed in many spiral galaxies and have long attracted attention due to their unique value in the study of galaxy evolution \citep{hoag1950peculiar,kormendy1979morphological,buta1996galactic}. Ringed galaxies are a subclass of spiral galaxies characterized by concentric ring structures surrounding their cores. These structures are often associated with specific dynamical mechanisms, including star-forming regions, density waves, and interactions between galaxies \citep{fosbury1977a0035,schweizer1983colliding,buta1990weakly,buta2017galactic,elmegreen1992influence,appleton1996collisional,grouchy2010ring,herrera2015catalogue,fernandez2021properties}. These ring structures provide a valuable window for understanding the internal dynamics and evolutionary history of galaxies \citep{lynds1976interpretation,theys1976ring,theys1977ring,schwarz1981response,struck1990varieties,romano2008stellar,elagali2018ring,murugeshan2023hi,tous2023origin}.

Rings in galaxies can primarily be divided into two types: normal rings (also known as resonance rings) and catastrophic rings. Normal rings are generally believed to form due to the continuous action of gravitational torques generated by bar structures, causing gas to accumulate at certain resonance points 
\citep{buta1996galactic}. In the absence of a bar structure, normal rings can also be driven by elliptical or spiral potential fields \citep{schwarz1981response}. Catastrophic rings, on the other hand, are typically considered the result of galaxy collisions, which can produce three types of catastrophic rings: accretion rings \citep{schweizer1987structure}, polar rings, and collisional rings (or R rings) \citep{lynds1976interpretation}. To explain the observed ring structures, scientists have proposed various theories, with resonance theory being the most widely accepted. Additionally, the "manifold theory" \citep{athanassoula2009ringsa,athanassoula2009ringsb,athanassoula2010rings} has made significant progress in describing the morphology of rings.

Although ring structures hold significant scientific value, successfully analyzing the characteristics of ringed galaxies still requires the support of a statistically meaningful large sample to fully reveal the underlying physical properties. In recent years, with the development of large-scale imaging survey projects such as the Sloan Digital Sky Survey (SDSS) \citep{fukugita1996sloan,york2000sloan}, the Legacy Survey of Space and Time (LSST) \citep{ivezic2019lsst,abell2009lsst}, the Euclid Space Telescope (Euclid) \citep{laureijs2011euclid}, the Kilo-Degree Survey (KiDS) \citep{de2013kilo}, the Dark Energy Survey (DES) \cite{collaboration2016more,abbott2021dark}, and the upcoming Chinese Space Station Telescope (CSST) \citep{zhan2021wide}, astronomers have obtained millions of high-resolution multi-band images. This creates an opportunity to construct comprehensive catalogs of ringed galaxies and conduct detailed studies \citep{buta2013galaxy}. However, the detection and identification of ring structures still face many challenges within these large-scale astronomical surveys.

On one hand, the vast data generated by modern telescopes raises higher demands on analysis methods and efficiency. Although traditional human visual classification methods have advantages in identifying major and minor galaxy morphologies, reliably distinguishing basic types and capturing subtle details that are difficult for computers to characterize \citep{buta2017galactic}, their efficiency is far lower than that of automated algorithms. On the other hand, ring features are often quite subtle and can be significantly affected by observational conditions. This not only imposes stricter requirements on the physical resolution of observational images but also challenges the performance of automated identification algorithms, thereby further increasing the difficulty of research. Therefore, developing efficient and reliable automated algorithms to accurately identify these structures has become an important direction in current astronomical research.

In recent years, with the rapid development of deep learning technology, data-driven methods for galaxy morphology recognition have gradually emerged, demonstrating enormous potential. The application of convolutional neural networks (CNNs), in particular, has significantly improved the accuracy of ringed galaxy identification \citep{banerji2010galaxy,dieleman2015rotation,dominguez2018improving,abraham2018detection,shimakawa2024galaxy,krishnakumar2024analysis,abraham2025automated}. Through convolutional layers, CNNs can effectively capture the spatial relationships and texture information within galaxy images, while automatically extracting potential patterns and features from large volumes of image data, enabling efficient and accurate morphology recognition. Furthermore, the application of deep learning technology not only significantly enhances processing efficiency but also reduces human bias during morphology recognition, making the results more objective and reliable.

However, the current machine learning models for identifying ringed galaxies are primarily built within a supervised learning framework, and their performance largely depends on the quality and quantity of labeled data. The annotation of galaxy rings is typically time-consuming and costly; for instance, \citet{buta1995catalog} spent nearly a decade compiling the catalog of Southern Ringed Galaxies (CSRG). Many standard datasets available for galaxy morphology classification, such as Galaxy10 DECals, do not label ringed galaxies as an independent category. Even in the Galaxy Zoo 2 (GZ2) dataset, which involved volunteer participation, only a small number of ringed galaxies were recorded, with at least 50\% consensus among volunteers. As a result, the scarcity of high-quality annotated data for ringed galaxies severely restricts the application of machine learning models in this area.

To overcome this bottleneck, researchers have begun to explore new methods. Data augmentation techniques are considered an effective solution to this problem. By transforming a small amount of high-quality labeled data, more annotated samples can be generated, effectively expanding the labeled dataset available for training and improving the morphology recognition performance of supervised learning models. For example, \citet{abraham2025automated} successfully trained a CNN using only 1,122 ringed galaxies from the SDSS catalog as training data. They augmented the training set through operations such as horizontal and vertical flipping, arbitrary angle rotation, and adjustments to brightness and contrast. They generated 4,855 high-quality ringed galaxy catalog entries from bright galaxy images (with extinction-corrected g-band magnitudes less than 16) in SDSS DR18. However, for some complex and diverse ring morphology features, relying solely on data augmentation may not be sufficient to fully uncover their intrinsic patterns, leading to insufficient generalization capability of the model in practical applications.

Another alternative to manual labeling is the use of synthetic data, which shows significant potential in reducing annotation costs. For example, \citet{krishnakumar2024analysis} pre-trained a CNN model using the Inception-ResNet V2 architecture \citep{szegedy2017inception} on 100,000 simulated galaxies, then fine-tuned the learned features on 3,117 real ringed galaxy samples before applying it to galaxy catalogs identified in the Pan-STARRS survey \citep{goddard2020catalog}. By incorporating Generative Adversarial Networks (GANs) for data augmentation, they successfully detected an additional 1,967 ringed galaxies. However, synthetic data typically represent simplified or simulated versions of real data, creating a distribution gap with actual observations. Consequently, modeling or inference based on synthetic data carries additional risks \citep{dagli2023astroformer}. \citet{krishnakumar2024analysis} noted that their model achieved only 58.9\% accuracy, posing challenges for large-scale imaging survey applications.

Recently, a technique called transfer learning \citep{zhuang2020comprehensive} has been applied to the identification of galaxy morphology. The core idea is to first train a deep learning model on a large-scale and well-annotated dataset, for example, using datasets that contain numerous galaxy images along with their morphological labels for pre-training. In this way, the model can learn the fundamental features and patterns associated with galaxy shapes. Subsequently, the pre-trained model is fine-tuned on a target dataset, which may contain fewer annotated samples but is more closely related to the specific research task at hand. This approach not only makes effective use of the knowledge learned from the pre-training phase but also reduces reliance on large amounts of labeled data while improving performance on specific tasks.

Researchers have discovered that the datasets used for pre-training can even be unrelated to galaxies, such as ImageNet \citep{russakovsky2015imagenet}. By making adaptive adjustments to the model’s head architecture for new tasks, the model can leverage the general features learned from pre-trained datasets and apply them to galaxy morphology recognition. For instance, \citet{shimakawa2024galaxy} utilized a deep CNN \citep{lecun1995convolutional,lecun2015deep}, which had been pre-trained on ImageNet, for classifying spiral and ring galaxies. They then adapted the model to the annotated dataset provided by GALAXY CRUISE \citep{tanaka2023galaxy} and performed fine-tuning. After this process, they applied the optimized model to public HSC-SSP datasets of nearby bright galaxy images (with a limiting magnitude of $r < 17.8$ and redshift ranging from $z = 0.01$ to 0.3), achieving promising results.

\citet{walmsley2022practical} found that models pretrained on all Galaxy Zoo DECaLS tasks, when fine-tuned with only a small number of newly labeled ring galaxies, perform significantly better in identifying ring galaxies compared to models pretrained on ImageNet or trained from scratch. They developed zoobot code, which enables effective model adaptation to new tasks even with limited annotated data. This approach not only substantially enhances model adaptability and flexibility but also drastically reduces reliance on large-scale labeled datasets, thereby making deep learning techniques more efficient and practical in astronomical research.

Nevertheless, the application of transfer learning faces several challenges. First, it is essential to evaluate the similarity between source and target tasks to ensure effective knowledge transfer; excessive disparity may lead to negative transfer, diminishing model performance. Second, selecting appropriate pretrained models and designing rational fine-tuning strategies are critical, including decisions on which layers to freeze, which to retrain, and how to adjust learning rates and other parameters. Lastly, discrepancies in data distributions between source and target domains can potentially impair the model’s generalization capabilities.

In this study, we adopted a novel deep learning architecture, GC-SWGAN, proposed by \citet{luo2025morphology}, for identifying ring structures in galaxy images from the DESI Legacy Imaging Survey. The GC-SWGAN model is based on a semi-supervised learning framework that effectively combines limited labeled datasets with abundant unlabeled datasets for training. This approach significantly reduces the reliance on large quantities of annotated data and provides an effective solution to technical challenges in identifying galaxy rings under low-resource conditions.

Semi-supervised learning has been successfully applied to various fields of astronomy \citep{huertas2023brief}, such as galaxy morphology classification \citep{wei2022unsupervised,ciprijanovic2023deepastrouda,vega2024nature,luo2025morphology}, black hole property estimation \citep{shen2021statistically}, detection of tidal features in galaxies\citep{desmons2024detecting}, radio galaxy classification \citep{slijepcevic2022learning,slijepcevic2024radio}, and solar magnetic field measurement classification \citep{lamdouar2022deep}. These applications demonstrate the effectiveness of semi-supervised learning in handling astronomical data.

Unlike methods based on transfer learning, the GC-SWGAN adpoted in this paper does not rely on pre-trained models or large-scale labeled databases. This characteristic makes it particularly suitable for addressing the scarcity of annotated data in astronomical research. By effectively utilizing the rich information contained in unlabeled data, GC-SWGAN maintains high level of morphological recognition accuracy and stability even under limited labeling conditions, thereby providing an efficient and practical solution for the dection of galactic rings. 

Additionally, after using the GC-SWGAN model to identify ringed galaxies within the DESI Legacy Imaging survey, this paper also analyzes the physical properties of the identified ringed galaxies, focusing on the differences in color and specific star formation rate (SSFR) distributions between ringed and non-ringed galaxies. These analyses help to gain a deeper understanding of the star formation characteristics of ringed galaxies and their differences compared to non-ringed galaxies.

The structure of this paper is organized as follows: Section 2 describes the data used in the study, detailing its characteristics and sources. Section 3 outlines the preprocessing procedures for the training data, including image cropping, normalization, augmentation, and the division into training and test sets. Section 4 presents the architecture of the semi-supervised GC-SWGAN neural network model used in this study, along with its training process. Section 5 discusses the training results of the GC-SWGAN model and its performance on the test set. Section 6 presents the catalog of ringed galaxies generated by applying our model to galaxy images from the DESI Legacy Imaging Surveys. Section 7 analyzes the properties of the ringed galaxies identified by the model, with a focus on comparing the differences in color and specific star formation rate (SSFR) distributions between ringed and non-ringed galaxies. Finally, Section 8 summarizes the research findings and discusses their implications.

\section{data} \label{sec:dataset}

Training a semi-supervised learning model for galaxy ring identification requires two types of data: one type consists of labeled galaxy samples, while the other consists of unlabeled galaxy samples. The labeled galaxy samples include both ring galaxies and non-ring galaxies. The quality and quantity of these labels are crucial and typically need to be carefully annotated by professional astronomers to ensure that the model can accurately learn the distinguishing features between ringed and non-ringed galaxies. The sources for the unlabeled galaxy samples are quite diverse, allowing for any number of galaxy images from the target domain to be selected for training.

Using observational images from ground-based telescopes to accurately identify ringed galaxies is an extremely challenging task \citep{shimakawa2022passive}. This challenge becomes particularly evident in the GZ2 dataset. Despite the involvement of a large number of volunteers, there remains significant uncertainty in identifying rings. Within this dataset, only a small number of ringed galaxy classifications have achieved at least 50\% agreement among volunteers, and the resulting identifications exhibit notable biases.

\citet{willett2013galaxy} point out that, when lacking detailed explanations about different ring types, GZ2 classifiers tend to focus more on identifying larger-scale outer rings. Given this, relying on experienced galaxy morphology experts for annotation is crucial to obtain high-quality ringed galaxy training samples and achieve desired model training outcomes. This approach can not only effectively reduce subjective bias but also significantly enhance the accuracy and consistency of identification results.

We selected two widely-used catalogs of ringed galaxies to construct our labeled training set: the Catalog of Southern Ringed Galaxies (CSRG) \citep{buta1995catalog} and the Galaxy Zoo 2 Catalog of Northern Ringed Galaxies (GZ2-CNRG) \citep{buta2017galactic}. The CSRG, compiled by Buta in 1995, contains 3,692 galaxies with declinations below -17°. The GZ2-CNRG sample was generated through crowdsourcing, comprising 3,962 galaxies initially identified by GZ2 volunteers and subsequently verified by expert \citet{buta2017galactic}.

After merging these two catalogs, we downloaded their $g$, $r$, and $z$-band images from the DESI Legacy Imaging Surveys (DESI-LS) website\footnote{\url{https://www.legacysurvey.org/}}. Following the removal of unobserved targets and poor-quality images (e.g., those affected by artifacts, blurring, or overexposed areas), we ultimately obtained a final training sample consisting of 5,173 ringed galaxy examples for our study. Among them, 1390 ringed galaxies are from CSRG, and 3783 ringed galaxies are from GZ2-CNRG.

DESI-LS primarily provide imaging for the Dark Energy Spectroscopic Instrument (DESI; \citealt{dey2019overview}), located on the 4-meter Mayall Telescope at Kitt Peak in the United States, to support its cosmological studies. The DESI-LS comprises three independent and collaborative surveys: DECaLS, BASS, and MzLS. BASS and MzLS collectively cover the northern sky from Kitt Peak in the United States. BASS captures $g$ and $r$ band images using the 2.3-meter Bok Telescope \citep{williams200490prime}, while MzLS obtains $z$ band images with the same 4-meter Mayall Telescope that hosts DESI itself \citep{dey2016mosaic3}. These two surveys are jointly referred to as BASS/MzLS. DECaLS, on the other hand, captures $g$, $r$, and $z$ band images of the southern sky using the 4-meter Blanco Telescope at the Cerro Tololo Inter-American Observatory in Chile  \citep{flaugher2015dark}. Together with BASS/MzLS, DECaLS provides DESI with a total of 14,000 square degrees of $g$, $r$, and $z$ band target imaging. Equally important is the data from the Dark Energy Survey (DES; \citealt{dark2016dark}), which used the same instrument as DECaLS (the DECam on the 4-meter Blanco Telescope). Consequently, the $g$, $r$, and $z$ band images captured by DES are also included in the data releases of the DESI-LS. These images encompass the entire 5,000 square degree region around the southern galactic pole. In this study, we utilized the Data Release 9 (DR9) version of the DESI-LS.

In addition to high quality ringed galaxy training samples, well-prepared non-ringed galaxy training samples are equally crucial. To ensure the reliability and representativeness of our data set, we used volunteer voting data in the Galaxy Zoo 2 (GZ2) project \citep{walmsley2022galaxy}. Specifically, volunteers were asked: 'Are there any of these rare features?' to determine if a galaxy possesses ring structures. Based on previous studies, when the proportion ($f$) of positive votes is less than 0.05, a galaxy can be reliably classified as non-ringed \citep{walmsley2022practical}.

In this study, we adopted an even stricter criterion by exclusively selecting galaxies with zero votes for ring features. From these, approximately 15,000 galaxies were randomly chosen to serve as our non-ringed galaxy training samples. To ensure consistency in our analysis, the corresponding $g$, $r$, and $z$ band images of these selected galaxies were downloaded from the DESI-LS database for further processing.

\begin{figure} 
        \includegraphics[width=\textwidth]{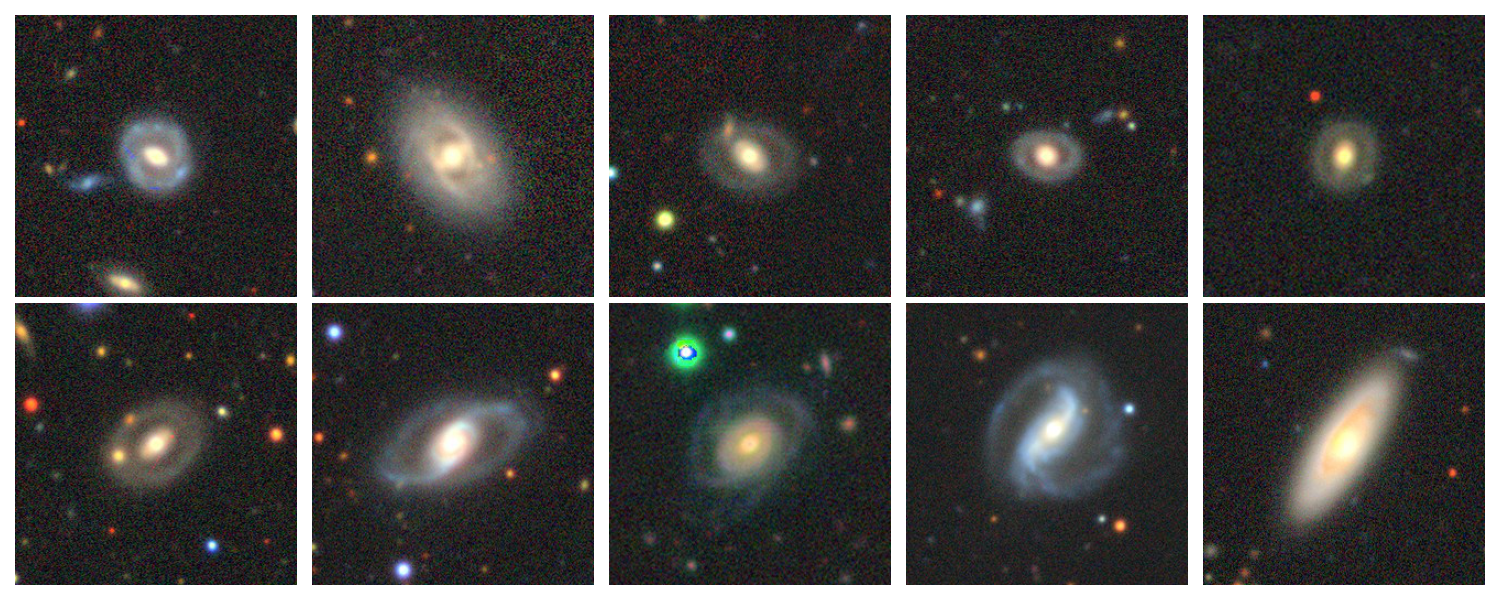} 
        \caption{Random samples of galaxies with rings from our training set.} 
        \label{fig:rings_train} 
\end{figure}

\begin{figure} 
        \includegraphics[width=\textwidth]{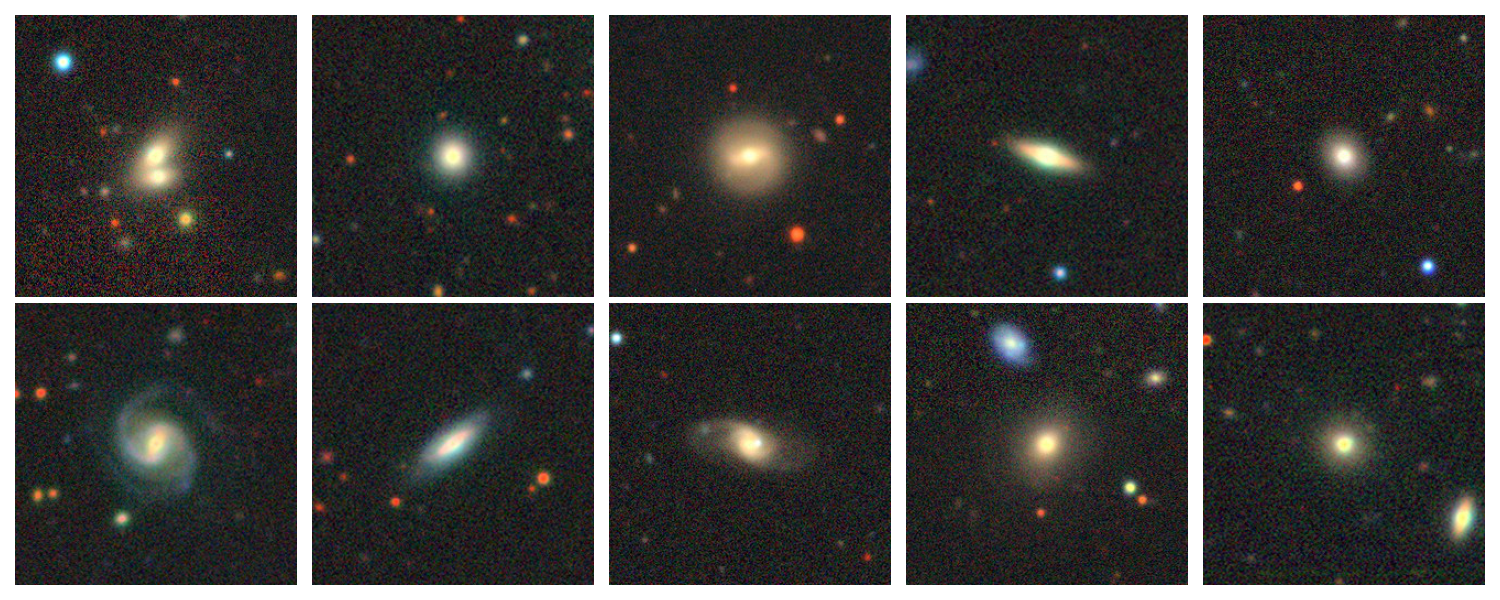} 
        \caption{Random samples of galaxies without rings from our training set.} 
        \label{fig:norings_train} 
\end{figure}

To ensure sample purity, one of our co-authors (JZ) performed careful visual inspection on hundreds of randomly selected images, confirming no misclassification cases in this non-ringed galaxy sample. Figures \ref{fig:rings_train} and \ref{fig:norings_train} present representative examples of ringed and non-ringed galaxies from our labeled training set, respectively.

For the unlabeled training samples, although the DESI-LS dataset provides a wealth of resources, in this study, due to limitations in computing hardware, we randomly selected approximately 15,000 galaxies from the Galaxy Zoo DESI catalog to serve as unlabeled samples for model training. Under better hardware conditions, a larger number of training samples may enhance model performance. Furthermore, we did not apply any specific selection criteria to filter the galaxy types or image quality within this subset of data. Firstly, our goal was to increase the diversity of the training data, which would improve the model’s generalization ability, enabling it to better recognize various types of galaxy features. Secondly, by including galaxy images of differing quality, we aimed to train a more robust model that can handle galaxy images under a range of observational conditions in practical applications.

Subsequently, we utilized the JPEG cropping tool \footnote{https://www.legacysurvey.org/viewer/} provided by the Legacy Survey website to download color images of the galaxies (in the $g$, $r$, and $z$ bands) from the DESI-LS website, based on the right ascension and declination coordinates of these unlabeled sample galaxies. These galaxy images cover a wide variety of types, ensuring that the model can learn rich morphological features from them.

It is worth noting that the DESI-LS website contains millions of unlabeled DESI galaxy images, offering great potential for expanding the unlabeled dataset. By further enlarging the unlabeled dataset, we can provide the model with a broader range of training resources, thereby enhancing its generalization capability and recognition accuracy.

\section{DATA PREPROCESSING} \label{sec:data_pre}

To enhance model performance and ensure consistency between labeled and unlabeled data, we implemented a unified preprocessing workflow for both types of datasets. The preprocessing steps primarily included image cropping, pixel value normalization, data augmentation, and dataset partitioning.

Firstly, regarding image cropping, we directly downloaded raw images from the DESI-LS website without any scaling or normalization of size. These images have a uniform size of 256×256 pixels, with each pixel corresponding to a resolution of 0.262 arcseconds. Additionally, the primary galaxies are positioned at the center of these images. To balance computational efficiency and model performance, we cropped these images from 256×256 pixels to 192×192 pixels. This strategy significantly reduces computational resource consumption by approximately 32\% while retaining the primary structural features of galaxies. The 192×192 pixel images correspond to an angular size of about 50×50 arcseconds for galaxies, which has been validated to adequately preserve the main structural characteristics for the vast majority of galaxies in our sample \citep{radford2015unsupervised,luo2025cross}. However, for a very small number of ringed galaxies with highly extended structures, there is still the possibility of losing important feature information. In future studies, we aim to explore the use of larger-sized images for analysis when resources permit.

Secondly, in terms of pixel normalization, the original image pixel values range from [0, 255]. We use the following formula to normalize them to a range of [-1, 1]:

\begin{equation}
     x^* = \frac{x - 127.5}{127.5},
     \label{eq:normal} 
\end{equation}
where $x$ represents the original value of each pixel, and 127.5 is the midpoint value of the [0, 255] range. The advantage of this normalization method lies in its ability to distribute the data symmetrically within a [-1, 1] interval, which facilitates gradient optimization during the model training process. Additionally, this approach preserves image details more effectively and accelerates the convergence speed of the model.

Next, regarding data augmentation, to enhance the model’s generalization capability, we employed three simple yet effective geometric transformation methods: horizontal flipping (random horizontal mirroring of the image), vertical flipping (random vertical mirroring of the image), and 90-degree rotation (rotating the image either clockwise or counterclockwise by 90 degrees). Research has shown that these transformations can effectively capture the symmetric features of galaxy morphologies while not significantly increasing computational costs \citep{dieleman2015rotation,kim2016star,cavanagh2021morphological}.

Lastly, regarding dataset partitioning, for labeled data, we allocated them into training and testing sets with a ratio of 8:2. The testing set comprises 1,028 ringed galaxies and 2,996 non-ringed galaxies. For unlabeled data, we utilized all 15,000 galaxy images mentioned in Section \ref{sec:dataset}. Given the abundant data resources of the DESI-LS, there is still significant potential to expand this dataset in the future.

\section{Methodology} \label{sec:method}

For galaxy morphology classification, supervised machine learning models typically rely on large amounts of labeled data to optimize complex free parameters, ensuring high accuracy and generalization capabilities. However, ringed galaxies are rare and challenging to identify, with only hundreds to thousands of samples available in existing expert catalogs. In this study, although we have collected over 5,000 ring galaxy samples reviewed by galaxy morphology experts, this dataset is still insufficient for training supervised learning models. To address the shortage of labeled data, we adopted the semi-supervised learning framework proposed by \citet{luo2025morphology}, named GC-SWGAN. This method combines Generative Adversarial Networks (GAN; \citealt{goodfellow2014generative}) with representation clustering techniques, enabling effective learning from a large amount of unlabeled data and significantly reducing reliance on labeled samples. For detailed information about GC-SWGAN, please refer to the work of \citet{luo2025morphology}. In this section, we will briefly outline the main principles of this method and its specific application in ringed galaxy identification tasks, as well as describe the model’s training process.

\subsection{GC-SWGAN Model}

GC-SWGAN is an innovative semi-supervised multi-task learning model that integrates the advantages of semi-supervised generative adversarial networks (SGAN; \citealt{odena2016semi}) and Wasserstein GANs with gradient penalty (WGAN-GP; \citealt{arjovsky2017wasserstein,gulrajani2017improved,adler2018banach}). This model achieves stable high-precision galaxy classification even when labeled data is limited \citep{luo2025morphology}. Specifically, the integration of SGAN effectively utilizes both scarce labeled and abundant unlabeled data to significantly enhance the model's learning capabilities. Meanwhile, the WGAN-GP architecture provides higher stability and convergence of training, allowing its widespread application in multiple fields \citep{gao2020data}, including astronomical studies \citep{rodriguez2018fast,kodi2020super,margalef2020detecting,li2021ai,rustige2023morphological,luo2025cross}. 

The model consists of three core components: the generator $G$, the discriminator $D$, and the classifier $C$. Although $D$ and $C$ are designed independently, they share a common feature space. The discriminator $D$ is responsible for distinguishing between real and generated data, while the classifier $C$ focuses on the classification task. Through collaborative optimization with the generator $G$, both components significantly enhance the overall performance of the model while also improving its convergence and stability.

In GC-SWGAN, the classifier $C$ is trained directly using labeled data and shares the feature extraction layer with the discriminator $D$. This design allows $C$ and $D$ to achieve information sharing in the feature space, further enhancing classification performance. Since the classifier and discriminator share part of the network structure, optimizing the discriminator indirectly optimizes the classifier as well. This unique collaborative optimization mechanism creates a dynamic feedback loop: improvements in the discriminator’s performance provide more accurate gradient feedback to the generator, facilitating the generation of more representative pseudo-samples; in turn, as the generator improves, adversarial training further enhances the discriminator’s feature extraction capabilities, indirectly optimizing the classifier’s classification performance. This progressive mechanism enables the model to achieve performance improvements in both data generation and classification tasks.

Experiments conducted by \citet{luo2025morphology} on the Galaxy10 DECaLs dataset have shown that, compared to traditional fully supervised methods, GC-SWGAN requires only about one-fifth of the labeled data to achieve comparable classification accuracy. This significant advantage makes it an ideal choice for galaxy morphology classification tasks in low-labeled data environments and demonstrates its outstanding performance in improving model generalization.

We have made appropriate modifications to the GC-SWGAN model to adapt it to the task of ringed galaxy identification in this study. Its network architecture, as shown in Figure 3, mainly consists of three core components: the generator $G$, the discriminator $D$, and the classifier $C$. The generator $G$ is composed of fully connected layers, a series of transposed convolution layers, batch normalization layers, and Leaky ReLU activation functions, with the output layer using a $tanh$ activation function to ensure that the pixel values of the generated images fall within the range of [-1, 1]. The discriminator $D$ and the classifier $C$ are designed independently but share part of the network structure, which includes multiple layers of CNNs, batch normalization layers, and Leaky $ReLU$ activation functions responsible for extracting multi-level features from the input data. The non-shared parts consist entirely of fully connected layers that fulfill the specific tasks of the discriminator and classifier, respectively. This shared design not only reduces the number of model parameters and improves computational efficiency but also enables the discriminator and classifier to collaboratively optimize during the feature extraction phase, further enhancing overall performance. For detailed network structures of $D$, $C$, and $G$, please refer to the work of \citet{luo2025morphology}. Moreover, the code utilized in this study is openly accessible on our GitHub repository \footnote{https://github.com/zjluo-code/GC-SWGAN-Rings} for further exploration and experimentation.

\begin{figure} 
        \includegraphics[width=\textwidth]{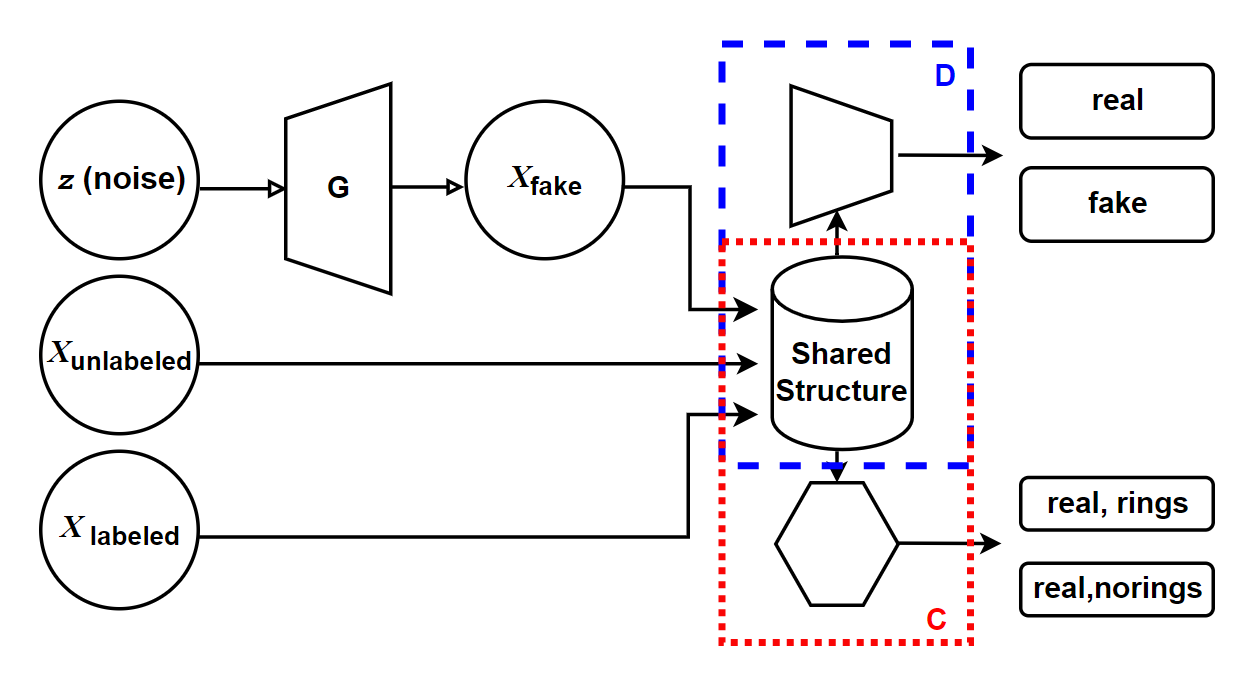} 
        \caption{The network architecture of GC-SWGAN. The symbol $z$ represents the input random noise vector used by the generator to produce synthetic images. $X_{unlabeled}$ denotes the input unlabeled real images, $X_{labeled}$ indicates the input labeled real images, and $X_{fake}$ represents the synthetic images generated by the generator $G$. The generator $G$ employs a multi-layer convolutional network structure, responsible for generating synthetic images that resemble real images. The task of the discriminator $D$ is to distinguish between real and fake images, differentiating real images from those generated by the generator. The classifier $C$ aims to learn how to assign correct class labels to real samples. In this study, the classifier $C$ focuses on differentiating real samples as ringed galaxies and non-ringed galaxies.} 
        \label{fig:framework} 
\end{figure}

To better accomplish the task of identifying ringed galaxies, we modified the last layer of the classifier $C$ by replacing the activation function of the output layer with one that is more suitable for binary classification tasks. Specifically, we adopted the $Sigmoid$ activation function to output the probability that a sample belongs to the ringed galaxy category. This modification enables the classifier to accurately handle the binary classification problem of whether galaxies have rings or not.

\subsection{Training the Model}

During the training process for the ringed galaxy identification task, the GC-SWGAN model needs to handle three different sources of data. The first type of data is the labeled real images $X_{labeled}$, which include the annotated samples we collected in Section 2: 5,173 ringed galaxies and 15,000 non-ringed galaxies. The discriminator $D$ will learn to distinguish between real samples and generated samples, while the classifier $C$ will learn to predict whether a sample is a ringed galaxy. The joint training of both components allows the model to simultaneously optimize its feature extraction capability and classification performance.

The second type of data is the unlabeled real images $X_{unlabeled}$. During the training process, only the discriminator $D$ is involved in processing this portion of the data, optimizing its feature extraction capability by assessing the authenticity of the samples. This study uses 15,000 unlabeled DESI-LS images collected in Section 2 as the source of unlabeled real data.

The third type of data utilized in our model consists of synthetic images denoted as $X_{fake}$, generated by the generator $G$ through the introduction of a random noise vector $z$. The discriminator $D$ plays a critical role in this process by distinguishing these synthetic images as counterfeit, thereby enhancing its ability to discern between real and fraudulent data. It is important to note that neither the second nor the third dataset directly participates in the training of the classifier $C$. However, due to the shared network architecture between the classifier $C$ and the discriminator $D$, optimizing the performance of $D$ leads to an indirect enhancement of $C$’s capabilities as well. 

The loss function of the GC-SWGAN model consists of three components: generator loss $L_G$, discriminator loss $L_D$, and classifier loss $L_C$. The generator loss $L_G$ is an adversarial loss aimed at maximizing the probability that the generated images are classified as real by the discriminator, and we calculate this using the Wasserstein distance. The discriminator loss $L_D$ is composed of two parts. The first part also uses the Wasserstein distance to measure the difference between real samples and generated samples. The second part is a gradient penalty term, which enforces Lipschitz continuity to avoid gradient explosion or vanishing. The classifier loss $L_C$ is a supervised loss, calculated using the categorical cross-entropy function. For a detailed computation of the model’s loss functions, please refer to the work of \citet{luo2025morphology}.

The model development employs the Keras API built on TensorFlow 2 \citep{abadi2016tensorflow}. The training process is conducted on a platform equipped with an NVIDIA L40S GPU. The batch size is set to 64. For optimization, all the generator, discriminator and classifier utilize the ADAM optimizer \citep{kingma2014adam}, configured with parameters $\beta_1 = 0.5$ and $\beta_2 = 0.999$. The initial learning rate is set to 0.0001, with exponential decay applied after each iteration at a decay factor of 1/1.000004. The training continues until it reaches 100,000 iterations and takes approximately 19.8 hours to complete.

\section{Model Performance Evaluation} \label{sec:performance}

To comprehensively evaluate the model’s performance in distinguishing between ringed and non-ringed galaxies, we utilize several metrics:  accuracy, precision, recall, and F1-score. These metrics provide insights into the model’s capability to recognize ringed galaxies from multiple perspectives.

Accuracy refers to the proportion of correctly classified samples out of the total number of samples, reflecting the overall correctness of the model’s classifications. Its calculation formula is as follows:

\begin{equation}
     \text{Accuracy} = \frac{\text{TP} + \text{TN}}{\text{TP} + \text{TN} + \text{FP} + \text{FN}},
     \label{eq:accuracy} 
\end{equation}
where TP represents the number of true positives, TN indicates the number of true negatives, FP denotes the number of false positives, and FN signifies the number of false negatives. Precision is the proportion of actual positive samples (ringed galaxies) among the samples predicted as positive (non-ringed galaxies) by the model, reflecting the reliability of the model’s predictions for ringed galaxies. The calculation formula is:

\begin{equation}
    \text{Precision} = \frac{\text{TP}}{\text{TP} + \text{FP}}.
     \label{eq:precision} 
\end{equation}
Recall is the proportion of samples that are correctly predicted as positive (ringed galaxies) relative to the actual positive samples, reflecting the model’s ability to recognize ringed galaxies. The calculation formula is:

\begin{equation}
    \text{Recall} = \frac{\text{TP}}{\text{TP} + \text{FN}}.
     \label{eq:recall} 
\end{equation}
The F1-score is the harmonic mean of precision and recall, taking both metrics into account and balancing the relationship between them. Its calculation formula is: 
\begin{equation}
 \text{F1-Score} = 2 \times \frac{\text{Precision} \times \text{Recall}}{\text{Precision} + \text{Recall}}.
 \label{eq:f1score} 
\end{equation}

In addition, the area under the receiver operating characteristic curve (AU-ROC) and the area under the precision-recall curve (AU-PRC) are two widely recognized evaluation metrics for assessing model performance. The AU-ROC is calculated by determining the area beneath the receiver operating characteristic  curve (ROC). The ROC illustrates the relationship between the true positive rate (TPR) and the false positive rate (FPR) at various threshold levels. The expressions for TPR and FPR are as follows:
\begin{equation}
 \text{TPR} = \frac{\text{TP}} {\text{TP} + \text{FN}},
 \label{eq:tpr} 
\end{equation}
\begin{equation}
 \text{FPR} = \frac{\text{FP}} {\text{TN} + \text{FP}}.
 \label{eq:fpr} 
\end{equation}
In an ideal model, the value of AU-ROC is 1, while for an ineffective model that makes random guesses, this value is 0.5. Similarly, AU-PRC is defined as the area under the precision-recall curve (PRC). As the model’s performance improves, the value of AU-PRC approaches 1.

When we set the probability threshold for predicting galaxies as ringed galaxies to 0.50, the trained model achieved approximately 97\% classification accuracy on the test set. Table \ref{tab:metrics} summarizes all classification performance metrics obtained by the model across the two categories of ringed galaxies and non-ringed galaxies. From the table, we can see that the model exhibits a considerable ability to intelligently identify ringed galaxies: on the test set, the precision for identifying ringed galaxies is approximately 94\%, the recall is about 93\%, the F1 score is 93\%, and both the AU-ROC and AU-PRC values exceed 0.97.

\begin{table}[ht]
    \centering
    \vspace{0.3cm} %
    \caption{Classification performance metrics of the model for ringed galaxies and non-ringed galaxies in the test set. \label{tab:metrics}}
    \setlength{\tabcolsep}{12pt} 
    \renewcommand{\arraystretch}{1.3} 
    \begin{tabular}{rccccc} %
        \hline
        \textbf{Class} & \textbf{Precision (\%)}  & \textbf{Recall (\%)}  & \textbf{F1-Score (\%)} & \textbf{AU-ROC} & \textbf{AU-PRC}  \\ 
        \hline
        Rings & 93.81 & 92.80 & 93.30 & 0.9921 & 0.9794\\
        Non-Rings & 97.50 & 97.86 & 97.68 & 0.9921 & 0.9970 \\ 
        \hline
    \end{tabular}
    \vspace{0.3cm} %
\end{table}

In our training samples, there is a certain degree of class imbalance between annotated ringed galaxies and non-ringed galaxies. To more accurately evaluate model performance, we further calculated the Matthews correlation coefficient (MCC) on the test set. As an evaluation metric that is better suited for imbalanced datasets, MCC is defined as follows:

\begin{equation}
 \text{MCC} = \frac{\text{TP} \times \text{TN} - \text{FP} \times \text{FN}}{\sqrt{(\text{TP} + \text{FP})(\text{TP} + \text{FN})(\text{TN} + \text{FP})(\text{TN} + \text{FN})}}.
 \label{eq:mcc} 
\end{equation}
The range of MCC is from -1 to +1, where +1 indicates perfect prediction, -1 indicates complete opposition to the true labels, and 0 indicates random prediction. In our experiments, the model achieved an MCC value of 0.91 on the test set, which demonstrates that even in the presence of class imbalance in the training samples, the model exhibits excellent performance in classifying ringed galaxies and non-ringed galaxies.

\begin{figure} 
        \includegraphics[width=\textwidth]{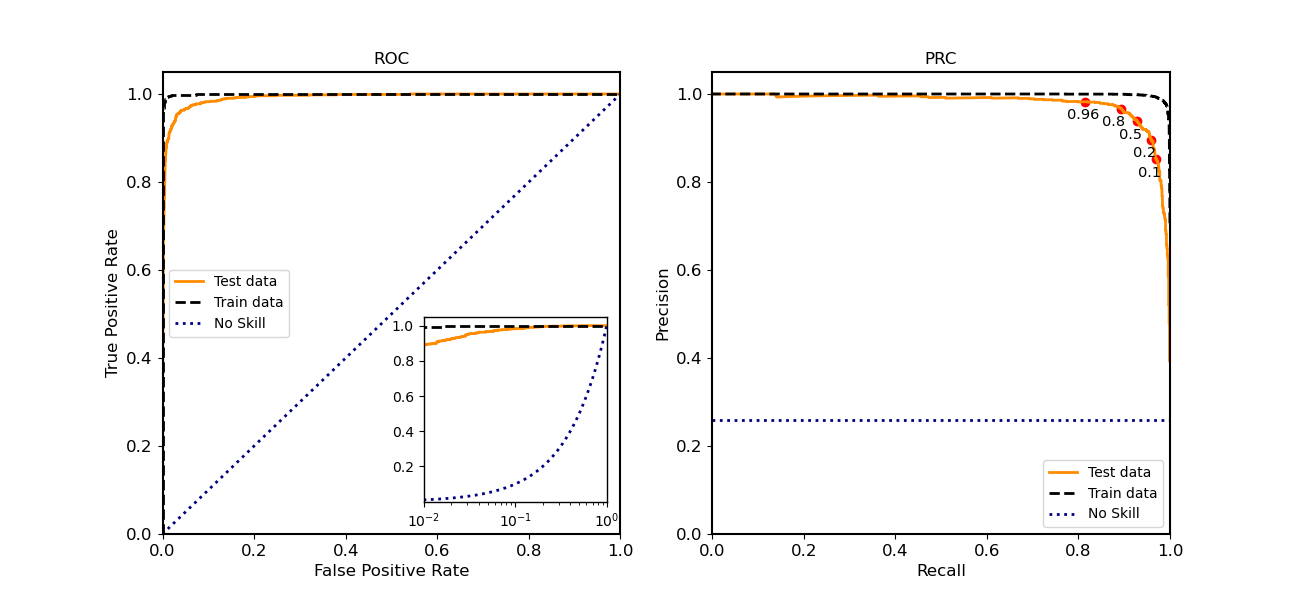} 
        \caption{Classification performance of the model for ringed galaxies. Left panel: The receiver operating characteristic curve (ROC) of the model on the training data and test data. The embedded subplot shows the x-axis on a logarithmic scale.} The black dashed line represents the ROC curve for the training set, while the orange solid line indicates the ROC curve for the test set. The dotted navy line represents the performance benchmark of a random classifier (AU-ROC = 0.5). Right panel: The precision-recall curve (PRC) of the model’s classifier. The meaning of the lines is consistent with that of the left panel. The values on the curve represent different classification thresholds for ringed galaxies. 
        \label{fig:ROC} 
\end{figure}

Through appropriate selection of classification probability thresholds, results with higher purity can be obtained. However, achieving higher purity comes at the cost of reduced recall. The nature of this trade-off can be understood by examining the precision-recall curve (PRC) shown in the right panel of Figure \ref{fig:ROC}. The values marked on the curve represent the classification thresholds for ringed galaxies, the black dashed line represents the PRC of the training data, the orange solid line represents the PRC of the test data, and the dotted horizontal line represents the PRC when classifications are made randomly. As can be seen from the figure, as the classification threshold increases, the model’s precision in identifying ringed galaxies also increases; however, the recall decreases simultaneously.

The left panel of Figure \ref{fig:ROC} presents the receiver operating characteristic curve (ROC) for the model’s classification of ringed galaxies. To better observe the performance details under low false positive rates, the embedded subplot within the panel uses a logarithmic scale on the x-axis. In this curve, the black dashed line represents the ROC for the training data, while the orange solid line denotes the ROC for the test data. The dotted diagonal line indicates the ROC when the model has no classification ability at all (i.e., AU-ROC = 0.5). From the figure, we can see that the model’s performance on the test data is close to that on the training data, indicating that the model has good generalization capability. Furthermore, the AU-ROC for the test data is significantly higher than 0.5, suggesting that the model has a strong ability to distinguish between ringed and non-ringed galaxies.

\section{A Catalog of Ringed Galaxies in DESI-LS} \label{subsec:catalog}

To establish a reliable catalog of ringed galaxies from DESI-LS, we selected locally bright galaxies for prediction, setting the redshift and magnitude constraints as follows: $0.0005 < z < 0.25$ and $r < 17.0$. These constraints are designed to exclude galaxies whose morphological details are difficult to detect through either visual or automated inspections, thereby ensuring the reliability and utility of morphological identification for the selected sample. Additionally, these constraints align with the selection criteria we adopted when preparing our training data. In the training samples, unlabeled data, as well as ringed and non-ringed galaxies, predominantly fall within this range. By establishing these limits, we can more effectively select samples of ringed galaxies, providing a solid foundation for subsequent analyses and research.

Galaxy Zoo DESI provides a catalog of 8.67 million galaxies in the DESI-LS (including DECaLS, MzLS, and BASS, as well as DES) through Zenodo \footnote{https://zenodo.org/records/8360385}. By applying filters based on redshift and magnitude, we obtained a subsample of 748,601 galaxies that satisfy the conditions of $r < 17.0$ and $0.0005 < z < 0.25$. We used Python’s $Concurrent$ module \footnote{https://docs.python.org/3/library/concurrent.futures.html} to batch download all these galaxy images from the DESI-LS website. Subsequently, we utilized the well-trained GC-SWGAN model to predict whether these galaxies exhibit ring structures.

When the probability threshold for ringed galaxies is set at 0.50, we obtained a total of 62,962 ringed galaxies, which account for 8.4\% of the total input samples to the model. To our knowledge, this represents the largest sample of ringed galaxies to date, with a preview of the sample catalog shown in Table 2. Based on the evaluation results from the training set, the identification precision for ringed galaxies at this threshold is approximately 94\%, and the recall rate is around 93\%. However, it is important to note that there is a significant imbalance in the sample sizes of ringed versus non-ringed galaxies in our training set, with a ratio of about 1:3, while the actual proportion of ringed galaxies in observations is even smaller. This bias in data distribution may lead the model to misclassify more ringed galaxies as non-ringed when faced with real observational data, thereby reducing the recall rate for ringed galaxies. 

Our prediction of the proportion of ringed galaxies is slightly lower than some previous studies. For example, \citet{walmsley2022practical} used the voting data from Galaxy Zoo DECaLS volunteers and combined it with machine learning predicted morphological vote scores to establish a relatively clean sample of ringed and non-ringed galaxies for model training. They identified a large number of galaxies with ringed and non-ringed structures from DECaLS images and found that about 12\% of galaxies may have ringed structures, of which about 90\% are considered to be true ringed galaxies. In addition, in the study by \citet{shimakawa2024galaxy}, they performed morphological classification on 59,854 galaxies (limited to $r < 17.8$ mag) based on the third public data release of the Hyper Suprime-Cam Subaru Strategic Program, with a redshift range of z = 0.01 to 0.3. By using a deep learning algorithm, they identified 15\% of the sample, namely 8808 ringed galaxies.

Our relatively low proportion of ringed galaxies may stem from the strict screening strategy in the training sample. To ensure data quality and classification accuracy, we utilized the CSRG and GZ2-CNRG samples, both of which underwent rigorous validation by morphological experts. This approach largely mitigates the possibility of incorrectly identifying ring structures. Although this strict screening method may result in a relatively lower proportion of ringed galaxies, it significantly enhances the accuracy and consistency of the model and reduces the risk of false positives.

\begin{figure} 
        \includegraphics[width=\textwidth]{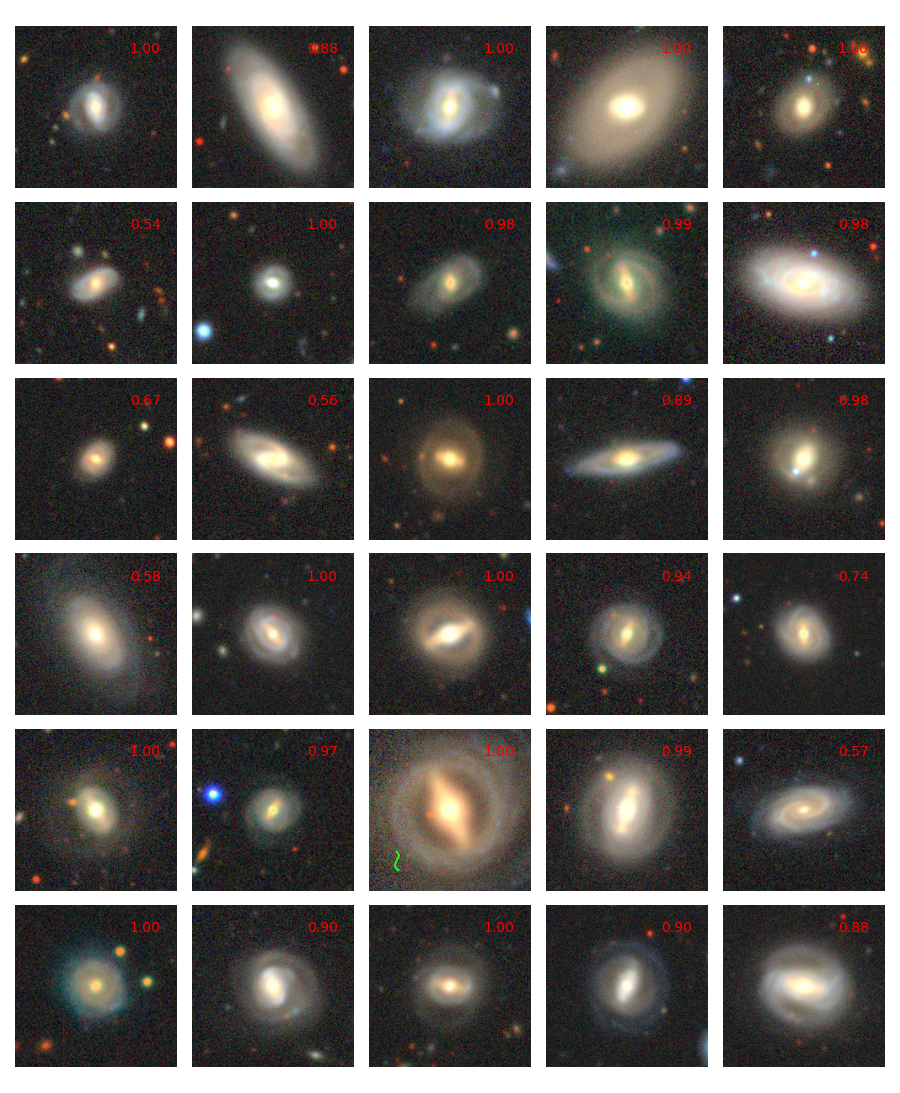} 
        \caption{ Images of 30 random galaxies from the catalog of 62,962 ringed galaxies detected by the model. The red numbers in each subplot indicate the probability of being predicted as a ringed galaxy.} 
        \label{fig:rings_desi} 
\end{figure}

Figure \ref{fig:rings_desi} showcases 30 randomly selected galaxies with ringed structures from our predicted catalog. As seen in the figure, the ringed structures of the vast majority of galaxies are quite apparent, which highlights the strong ring identification capability of our model. However, we also note that in a small number of samples, the ringed structures are somewhat ambiguous or exhibit other confounding features, indicating that there is still room for improvement in the model’s handling of complex or edge cases.

\begin{figure} 
        \centering
        \includegraphics[width=0.85\textwidth]{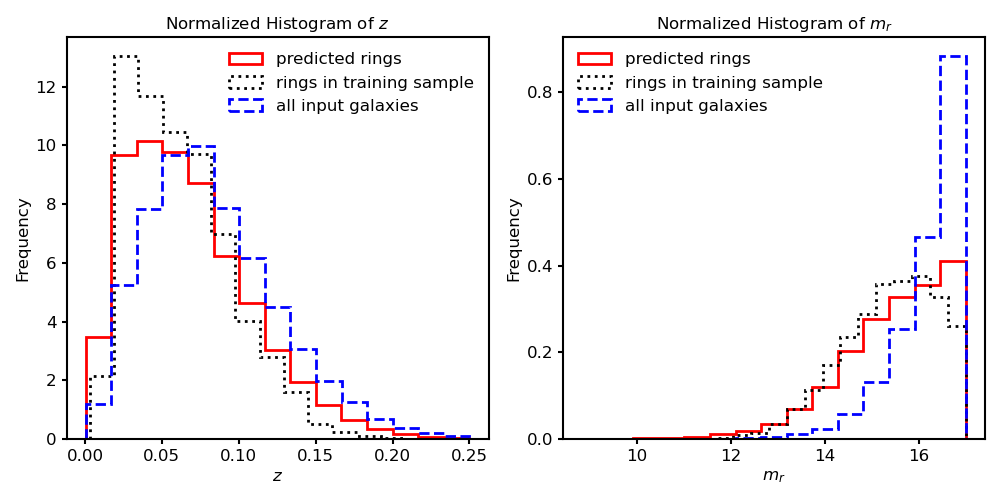} 
        \caption{Frequency histograms of three samples (input total galaxies, ring galaxies identified by the model, and ring galaxies used for training) for galaxy redshift $z$ (left panel) and DESI $r$-band apparent magnitude $m_r$ (right panel). The solid red line represents model-predicted ring galaxies, the dashed blue line represents all input galaxies, and the dotted black line represents ringed galaxies used in model training.} 
        \label{fig:z_mr_dist} 
\end{figure}

In addition, Figure \ref{fig:z_mr_dist} presents the distribution histograms of the ringed galaxy samples predicted by our model (62,962 galaxies) compared to the total input galaxy sample (748,601 galaxies) in terms of redshift $z$ and apparent magnitude $m_r$ in the $r$-band. In the figure, the red solid line represents the predicted ringed galaxy sample, while the blue dashed line represents the total input galaxy sample. It is evident from the figure that there are some differences in the distributions of these two samples in redshift and apparent magnitude in the $r$-band. Specifically, the predicted ringed galaxy sample tends to be concentrated at lower redshifts and higher luminosities. This result aligns with our expectations, as it is easier to identify ringed structures under conditions of low redshift and high brightness. Low redshift indicates that the galaxies are relatively close, allowing for clearer observation of structural details, which facilitates the detection of ringed features. Additionally, high-luminosity galaxies typically possess more pronounced morphological characteristics, making it easier to recognize ringed structures in images. 

\begin{figure} 
        \centering
        \includegraphics[width=0.85\textwidth]{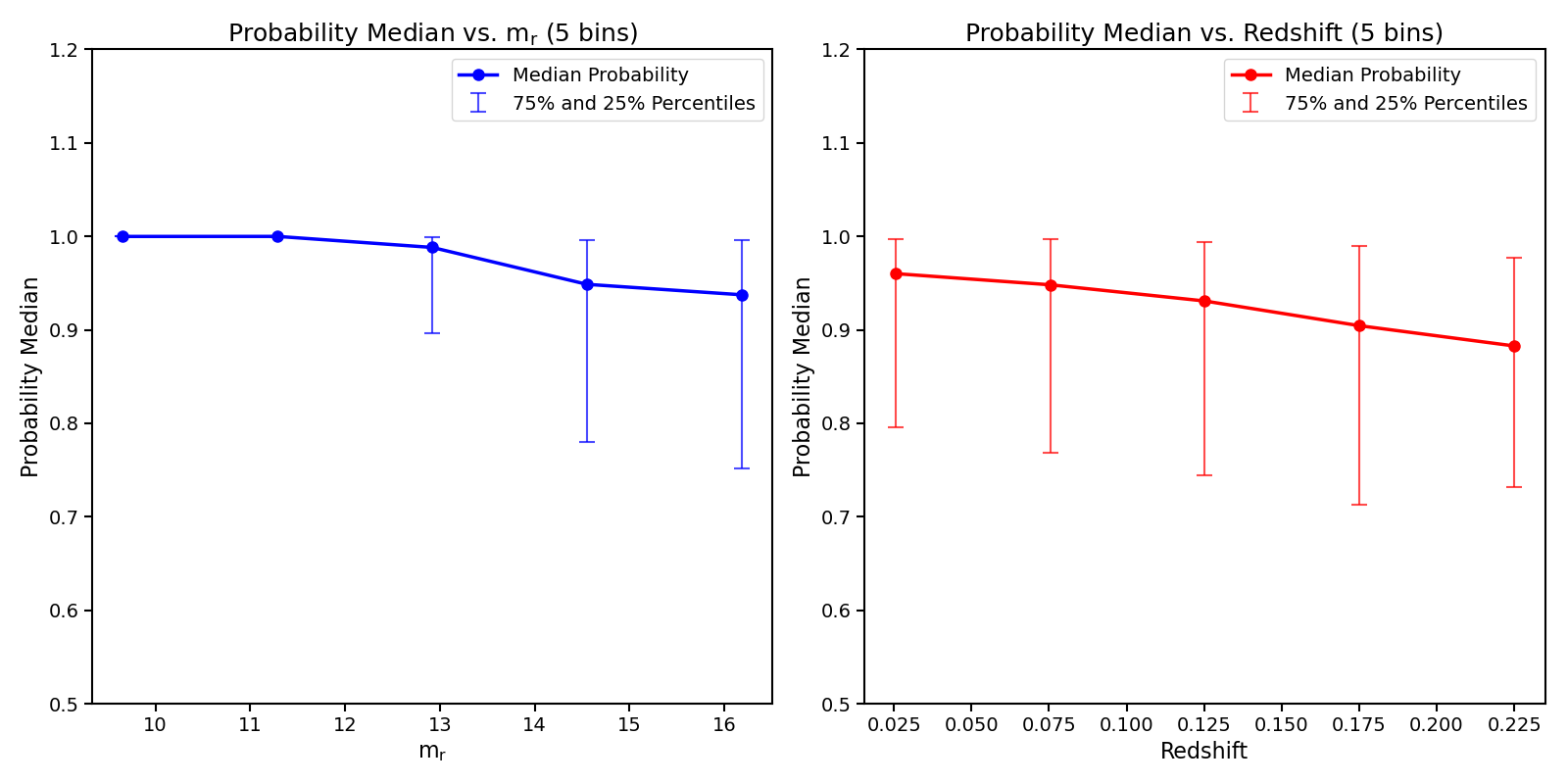} 
        \caption{Relationship between the probability of the model classifying a galaxy as a ringed galaxy and apparent magnitude $m_r$ (left panel) and redshift (right panel). Each point represents the median probability within the corresponding magnitude interval (left panel) or redshift interval (right panel), with error bars indicating the range from the 25th to the 75th percentile.} 
        \label{fig:prob_vs_mrz} 
\end{figure}

Further analysis is presented in Figure \ref{fig:prob_vs_mrz}, which illustrates the relationship between the probability of the model classifying a galaxy as a ringed galaxy and the apparent magnitude $m_r$ (left panel) and redshift (right panel). Each point in the figure represents the median probability within the corresponding magnitude range (left panel) or redshift range (right panel), while the error bars indicate the range between the 25th and 75th percentiles. The trend shown in the figure reveals that as $m_r$ decreases or redshift increases, the predicted probability of a galaxy being classified as a ringed galaxy also decreases, with the uncertainty gradually increasing.

Furthermore, the samples of ringed galaxies in our model’s training data are also primarily concentrated in the low redshift and high brightness range (as seen in the black dotted line in Figure 6), which may contribute to the model performing better under these conditions. Thus, the identified ringed structures are more likely to be found in the low redshift and high brightness regions, reflecting not only the limitations of observational conditions but also the distribution characteristics of the model’s training data.

\begin{figure} 
        \includegraphics[width=\textwidth]{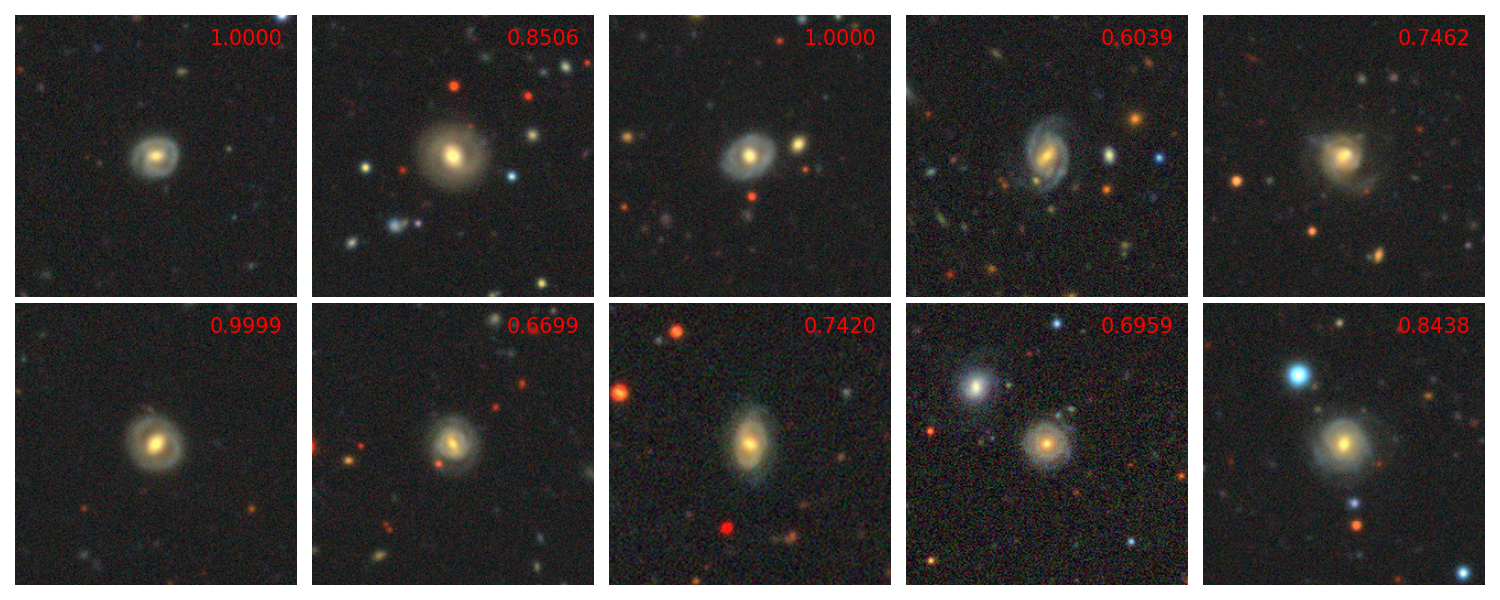} 
        \caption{ Images of 10 random galaxies predicted by the model to have ring structures. All these galaxies have $0.20 < z < 0.25$ and $16.5 < m_r < 17$.  The red numbers in each subplot indicate the probability of being predicted as a ringed galaxy.} 
        \label{fig:highz_lowlum} 
\end{figure}

To gain an intuitive understanding of the model’s performance in high redshift and low brightness ranges, we randomly selected 10 galaxies from the input sample that are located in the high redshift and low brightness interval ($0.20 < z < 0.25$, $16.5 < m_r < 17$) and were predicted by the model to have ring structures. Figure \ref{fig:highz_lowlum} shows the morphology of these galaxies. From the figure, it is evident that although the ring features of these galaxies are not as pronounced as those of galaxies at low redshift and high brightness, the ring structures of most galaxies can still be visually recognized. This result further confirms the strong capability of our model in identifying ring structures, indicating that it can effectively recognize ring structures even under more challenging observational conditions.

\begin{table}
    \centering
    \caption{Catalog of ring galaxies in DESI. 
    \label{tab:catalog}}
    \setlength{\tabcolsep}{10pt} 
    \renewcommand{\arraystretch}{1.06} 
    \begin{tabular}{cccccccc}
        \hline
        \textbf{dr8\_id} & \textbf{ra}  & \textbf{dec}  & \textbf{mag\_r\_desi} & \textbf{mag\_g\_desi} & \textbf{mag\_z\_desi} & \textbf{redshift} & \textbf{prob}  \\ 
        \hline
200045\_6104	& 11.210690 & -23.159163 & 15.125350 & 16.055828 & 14.440767 & 0.061653 & 0.997252 \\
201363\_3255 &	8.963280 &	-23.006286 &	15.003571 &	15.889303 &	14.319273 &	0.061062 &	0.929614  \\
200044\_2096 &	11.122921 &	-23.297175 &	14.764104 &	15.454452 &	14.227168 &	0.060218 &	0.893542 \\ 
204022\_350	& 9.306914 &	-22.586702 &	12.888265 &	13.512125 &	12.302115 &	0.010164 &	0.915831 \\
204023\_3423 &	9.539052 &	-22.504294 &	15.381775 &	16.138826 &	14.734895 &	0.090022 &	0.565160 \\
202696\_6361 &	10.383586 &	-22.643752 &	15.122063 &	15.845529 &	14.606272 &	0.063000 &	0.691074 \\
198721\_1605 &	10.720063 &	-23.540989 &	13.257116 &	14.047126 &	12.623620 &	0.022229 &	0.999984 \\
204022\_1575 &	9.393127 &	-22.549258 &	13.185019 &	14.042454 &	12.499637 &	0.012792 &	0.998959 \\
201363\_2826 &	9.096120 &	-23.016850 &	14.399699 &	15.226388 &	13.766101 &	0.064431 &	0.991956 \\
198721\_1606 &	10.724067 &	-23.545423 &	13.994370 &	14.751829 &	13.402916 &	0.022676 &	0.766204 \\
202697\_4988 &	10.648364 &	-22.687798 &	15.668744 &	16.253292 &	15.229361 &	0.050715 &	0.589824 \\
204068\_5883 &	21.788411 &	-22.444374 &	14.264937 &	15.045927 &	13.675238 &	0.032573 &	0.992424 \\
202696\_5334 &	10.430571 &	-22.682988 &	15.428199 &	16.373333 &	14.712164 &	0.063040 &	0.999788 \\
204022\_2892 &	9.248444 &	-22.514370 &	15.112688 &	15.975642 &	14.436160 &	0.064391 &	0.999584 \\
204069\_7609 &	22.060515 &	-22.375069 &	15.381672 &	16.095758 &	14.886155 &	0.031825 &	0.998862 \\
205353\_806 &	8.912720 &	-22.342475 &	15.738594 &	16.401297 &	15.194537 &	0.080926 &	0.672844 \\
206692\_1038 &	10.242701 &	-22.083076 &	14.083241 &	14.889816 &	13.437136 &	0.063671 &	0.986000 \\
208031\_2644 &	10.568728 &	-21.773967 &	14.925154 &	15.829144 &	14.263549 &	0.064795 &	0.845335 \\
208032\_5641 &	10.916096 &	-21.659210 &	15.222617 &	16.023643 &	14.588319 &	0.061823 &	0.943845 \\
208031\_1147 &	10.569286 &	-21.826379 &	15.316677 &	16.108229 &	14.693509 &	0.064228 &	0.999999 \\
209362\_2734	& 8.136038 &	-21.471848 &	15.159886 &	15.988932 &	14.492733 &	0.089502 &	0.570964 \\
209363\_3976	& 8.556886 &	-21.438498 &	12.914690 &	13.503025 &	12.379335 &	0.026845 &	0.528107 \\
209371\_4949	& 10.630500 &	-21.431160 &	15.563628 &	16.479858 &	14.885358 &	0.063154 &	0.951267 \\
204024\_3805	& 9.993092 &	-22.490721 &	14.856886 &	15.512996 &	14.345409 &	0.052583 &	0.866378 \\
212051\_5350	& 8.955859 &	-20.910507 &	15.268041 &	16.202920 &	14.588581 &	0.073007 &	0.998043 \\
212056\_1846	& 10.363190 &	-21.045335 &	12.177903 &	12.620642 &	11.831751 &	0.005190 &	0.998259 \\
214800\_4396	& 23.681303 &	-20.463979 &	13.787085 &	14.418366 &	13.328474 &	0.017792 &	0.999976 \\
210712\_6570	& 10.298942 &	-21.131586 &	14.803455 &	15.029070 &	14.710476 &	0.005517 &	0.996372 \\
214793\_5482	& 21.791369 &	-20.404246 &	13.626535 &	14.463169 &	12.989524 &	0.033523 &	0.999771 \\
206691\_4840	& 10.195278 &	-21.926008 &	14.911242 &	15.737783 &	14.293225 &	0.065002 &	0.997364 \\
217451\_3296	& 9.339773 &	-19.934236 &	11.790057 &	12.552162 &	11.206908 &	0.013042 &	1.000000 \\
216095\_4273	& 8.914493 &	-20.125388 &	12.928803 &	13.631742 &	12.387019 &	0.010984 &	0.956480 \\
216154\_3168	& 24.672911 &	-20.249725 &	13.976688 &	14.945221 &	13.166826 &	0.032336 &	0.976151 \\
218857\_3979	& 22.746126 &	-19.736532 &	15.033752 &	15.946837 &	14.365765 &	0.074315 &	0.886723 \\
221578\_1119	& 23.263347 &	-19.334607 &	15.009222 &	15.846705 &	14.378160 &	0.050815 &	0.977991 \\
218878\_4294	& 28.189927 &	-19.731716 &	14.760556 &	15.334837 &	14.366635 &	0.031935 &	0.987541 \\
222980\_6384	& 33.965365 &	-18.914327 &	13.951703 &	14.854547 &	13.251310 &	0.034424 &	0.999816 \\
224322\_2631	& 28.166080 &	-18.781392 &	13.793391 &	14.600221 &	13.168590 &	0.048634 &	0.999921 \\
224322\_2960	& 28.122065 &	-18.772261 &	13.995830 &	14.511896 &	13.637772 &	0.019137 &	0.980919 \\
224342\_976 	& 33.358330 &	-18.833318 &	14.302827 &	15.061026 &	13.685466 &	0.034731 &	0.986563 \\
224322\_519	    & 28.044552 &	-18.853126 &	15.016030 &	15.853209 &	14.404704 &	0.040094 &	0.999814 \\
224323\_4444	& 28.379827 &	-18.730871 &	15.061995 &	15.920317 &	14.454668 &	0.048037 &	0.818599 \\
224349\_6386	& 35.189213 &	-18.641698 &	14.546801 &	15.407008 &	13.864645 &	0.054077 &	0.999997 \\

\hline
\multicolumn{6}{l}{Note: The full version of this table is available online.}
\end{tabular}
\end{table}

To obtain a catalog with higher confidence, we can increase the classification threshold for ringed galaxies to 0.90, resulting in a total of 37,508 ringed galaxies identified. During the evaluation on the training set, this threshold corresponds to an accuracy of approximately 98\% for identifying ring structures. However, it is important to note that this accuracy is based on the training set and may not reflect the true performance of the model without further validation using independent datasets and true class labels. Furthermore, our model completes the classification of about 750,000 galaxies in less than 15 minutes on our GPU platform. This efficiency far surpasses the time taken by GZ2 volunteers, who spent 14 months classifying approximately 300,000 galaxies in SDSS DR7 and is significantly better than the supervised CNN model developed by \citet{krishnakumar2024analysis}. Their model with an Inception-ResNet V2 architecture took about 10 hours to classify roughly 950,000 galaxies.

\section{Properties of Ringed Galaxies}

The catalog of ringed galaxies generated by our model significantly expands the existing dataset for the study of ringed galaxies, providing a rich sample for in-depth analysis of their morphological features, formation mechanisms, and evolutionary paths. To demonstrate the practical value of the catalog produced in this study, we will focus on analyzing the color properties and star formation activities of ringed galaxies in this section. Specifically, by comparing differences between ringed and non-ringed galaxies in their optical color distribution ($g$-$r$) and specific star formation rate (SSFR, $\text{SFR}/M_*$), we aim to reveal the characteristics of galaxies with ring structures in terms of star formation behavior. Such comparative analyses will help us to understand the position of ringed galaxies within the entire star-forming sequence, as well as their relationships with the green valley galaxy population and the quiescent galaxy population.

\begin{figure} 
        \centering
        \includegraphics[width=0.85\textwidth]{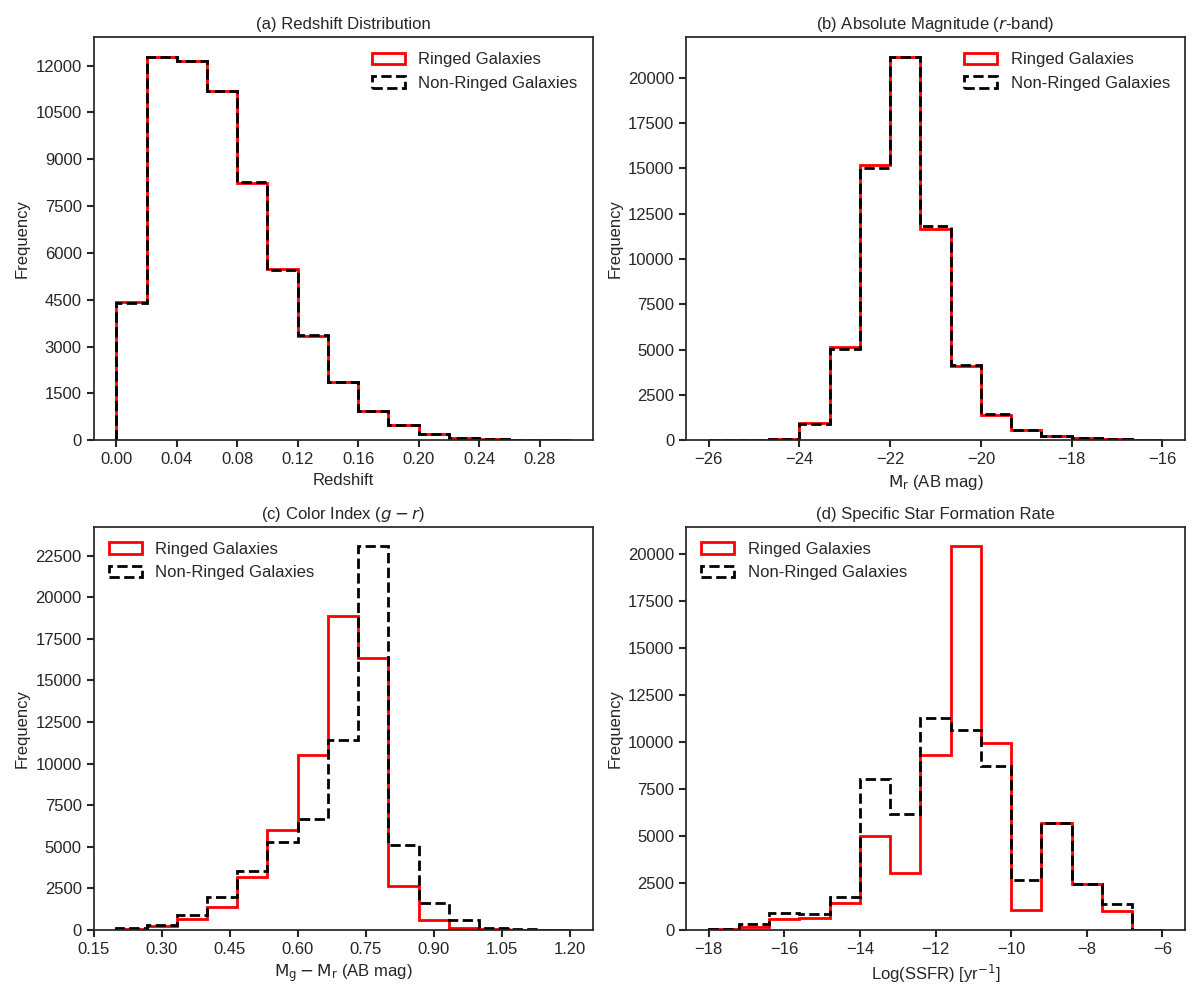} 
        \caption{Distributions of redshift, $\mathrm{M_r}$, $\mathrm{M_g-M_r}$, and $\mathrm{Log(SSFR)}$ for ringed galaxies (solid red lines) and for the control samples (non-ringed galaxies; dash black lines). Panels (a) and (b) demonstrate that the control galaxy sample exhibits high similarity to the ringed galaxy sample in terms of redshift distribution and $r$-band absolute magnitude ($\mathrm{M_r}$). In contrast, panels (c) and (d) reveal significant differences between the control samples and ringed galaxies in distributions of $g-r$ color and specific star formation rate (SSFR).}
        \label{fig:comp_color_sfr} 
\end{figure}

We constructed a control sample to reliably evaluate the differences between ringed and non-ringed galaxies by following the methods described in \citet{perez2009building} and \citet{fernandez2021properties}. Specifically, we used the Monte Carlo algorithm to select non-ringed galaxies whose redshift and $r$-band absolute magnitude ($\mathrm{M_r}$) distributions closely resemble those of the ringed galaxy sample. This process generated a control catalog, as shown in panels (a) and (b) of Figure \ref{fig:comp_color_sfr}.

Through these constraints, we obtained a control sample that matches the size of our ringed galaxy sample and exhibits highly similar distributions in both redshift and $r$-band luminosity. To validate the similarity between the two samples, we conducted a Kolmogorov-Smirnov (KS) test on their $\mathrm{M_r}$ and redshift distributions. The resulting $p$-values all exceeded 0.05, indicating that these samples likely originate from the same parent population. Since galaxy color and star formation properties are closely tied to stellar mass (luminosity) and redshift \citep{kauffmann2003dependence,kauffmann2003stellar,luo2007differences}, these constraints ensure comparability between the samples and provide a solid foundation for subsequent comparative analyses.

In our comparative analysis, we utilized the absolute magnitudes of the $g$-band and the $r$-band, as well as the specific star formation rate ($\text{SFR}/M_*$), for all galaxies. These data were extracted from the Galaxy Zoo DESI galaxy catalog \citep{walmsley2023galaxy} and were estimated based on the photometric redshift table from the DESI survey. Panels (c) and (d) of Figure \ref{fig:comp_color_sfr} compare the distributions of $g-r$ color and SSFR between our model identified ringed galaxies and the control sample.

From panel (c) in figure \ref{fig:comp_color_sfr}, we observe that the $g$-$r$ color distribution of ringed galaxies is more concentrated in the intermediate region compared to the control sample. This intermediate area, known as the 'green valley', represents the transition zone between the active and quiescent star formation regions. The corresponding values $\log(\text{SFR}/M_*)$ fall in the range of -11.8 to -10.8 \citep{salim2014green}. Furthermore, as shown in Figure \ref{fig:comp_color_sfr} panel (d), we find a marked increase in the number of ringed galaxies compared to the control sample within this $\log(\text{SFR}/M_*)$ range. These distribution characteristics indicate that more ringed galaxies are in this critical transition phase. Ring structures can significantly alter the properties of their host galaxies and may be associated with the complex dynamic changes in their internal star-forming activities.

\citet{fernandez2021properties,fernandez2024revealing} employed a similar approach to compare differences in color and specific star formation rate (SSFR) between ringed galaxies and non-ringed galaxies. In their studies, the criteria for constructing control samples differed from ours. They selected a non-ringed galaxy control sample that matched the ringed galaxy sample not only in redshift and magnitude but also in morphology and local density environment distributions, resulting in a control sample with a concentration index ($C$) similar to that of the ringed galaxies. Since ringed galaxies typically exhibit lower $C$ values, their control sample included a larger proportion of morphologically diffuse galaxies such as spiral and irregular galaxies. These types of galaxies are often gas-rich and exhibit more active star formation. Therefore, in their studies, ringed galaxies showed redder $g-r$ colors and lower SSFR levels compared to the non-ringed control samples.

In contrast, when constructing our control sample, we did not impose restrictions on the concentration index ($C$), resulting in a control sample that encompasses a broader range of morphological types, including elliptical galaxies. This difference in how the control sample was selected led us to observe that, overall, the ringed galaxies in our study exhibit bluer $g-r$ colors and slightly higher SSFR values compared to the non-ringed control counterpart. Despite these differences, our findings are consistent with those of \citet{fernandez2021properties,fernandez2024revealing}, both showing that ringed galaxies are more likely to reside in the green valley compared to non-ringed galaxies.

\section{SUMMARY}

In this study, we propose using a semi-supervised deep learning model GC-SWGAN to identify ringed structures within the DESI Legacy Imaging Survey. This approach aims to address the challenge of insufficient training samples commonly encountered by traditional supervised learning methods in identifying ringed galaxies.

To build a reliable foundational training dataset, we utilized the Catalog of Southern Ringed Galaxies (CSRG) and the Northern Ringed Galaxies from the GZ2 catalog (GZ2-CNRG) as annotated data for galaxies with rings. Furthermore, we extracted galaxies without rings from the Galaxy Zoo 2 dataset through strict screening. Additionally, we obtained unlabeled data from the Galaxy Zoo DESI catalog. Through a standardized image preprocessing workflow, including cropping, normalization, and enhancement steps, along with reasonable allocation of training and testing datasets, we provided high-quality input data for model training.

The GC-SWGAN model employed in this study integrates a semi-supervised generative adversarial network (SGAN) and Wasserstein GAN with Gradient Penalty (WGAN-GP). The discriminator and classifier in this model are independently designed yet share a feature space, optimizing performance through collaborative training with the generator. For the specific task of ringed galaxy identification, we adjusted the activation function of the classifier's final layer. Training was conducted on a computing platform equipped with an NVIDIA L40S GPU using the Keras API based on the TensorFlow framework. By appropriately configuring batch sizes, optimization algorithm parameters, and learning rate decay strategies, we ultimately achieved optimal training results.

The training results demonstrate the model's exceptional classification performance. On the test set, with a prediction probability threshold for ringed galaxies set at 0.5, the classification accuracy reached 97\%, while the precision and recall for ringed galaxy identification were 94\% and 93\%, respectively, yielding an F1-score of 93\%. Furthermore, both the AU-ROC and AU-PRC exceeded 0.97, indicating that the model excels at distinguishing between ringed and non-ringed galaxies while maintaining strong generalization capabilities.

Based on the well-trained model, we conducted a search for ringed galaxies in the DESI Legacy Imaging Survey. In order to obtain reliable samples of ringed galaxies, we further set reasonable input galaxy selection criteria: the redshift range was limited to $z \in (0.0005, 0.25)$, and the magnitude satisfied $m_r < 17.0$. Under these conditions, with a probability threshold of 0.5 for classification, we identified 62,962 ringed galaxy candidates from 748,601 galaxy images that met the requirements. This result significantly expanded the number of known ringed galaxies and built the largest ringed galaxy catalog to date. Through further analysis of the color and star formation properties of the ringed galaxies in this catalog, we found that ringed galaxies are more likely to reside in the “Green Valley” compared to non-ringed galaxies. This finding indicates that ring structures can significantly alter the properties of their host galaxies and may be associated with the complex changes in the star-forming processes within the galaxies. By further increasing the classification threshold to 0.9, while reducing the target count to 37,508, we saw a significant improvement in precision metrics. Additionally, the application results demonstrated that this model outperformed both Galaxy Zoo 2 volunteer annotations and \citet{krishnakumar2024analysis}’s supervised learning method in terms of classification efficiency. This highlights the enormous potential of semi-supervised deep learning methods for processing large-scale astronomical data.

However, this study also has some limitations. First, using images from ground-based telescopes for ringed galaxy identification presents certain challenges \citep{shimakawa2022passive}. The limitations in the quality of ground-based images may affect the model’s accurate identification of ringed galaxy features, thereby reducing the reliability of the classification. In addition, the ringed galaxies in the training set are primarily focused on low-redshift and high-luminosity galaxies, which limits the model’s practical identification capability for high-redshift and low-luminosity galaxies.

To overcome these limitations, future research could explore more advanced image processing techniques to improve image quality \citep{luo2025cross} and expand the training dataset to include a broader range of redshifts and luminosities. Additionally, it is recommended to utilize high-resolution images from space telescopes, such as the Euclid Wide Field Survey and the Chinese Space Station Telescope (CSST), to supplement data from ground-based telescopes, thereby enhancing the model’s generalization ability and identification accuracy, and providing essential foundational data for subsequent related research.

\begin{acknowledgments}
Z.J.L. acknowledges the support from the Shanghai Science and Technology Foundation Fund (Grant No. 20070502400) and the scientific research grants from the China Manned Space Project with Grand No. CMS-CSST-2025-A05 and CMS-CSST-2025-A07. 
C.C acknowledges support from the National Natural Science Foundation of China (Grant No. 12173045).
S.H.Z. acknowledges support from the National Natural Science Foundation of China (Grant No. 12173026), the National Key Research and Development Program of China (Grant No. 2022YFC2807303), the Shanghai Science and Technology Fund (Grant No. 23010503900), the Program for Professor of Special Appointment (Eastern Scholar) at Shanghai Institutions of Higher Learning, and the Shuguang Program (23SG39) of the Shanghai Education Development Foundation and Shanghai Municipal Education Commission. This work is also supported by the National Natural Science Foundation of China under Grant No. 12141302. Additionally, this work is sponsored (in part) by the Chinese Academy of Sciences (CAS), through a grant to the CAS South America Center for Astronomy (CASSACA).

The DESI Legacy Imaging Surveys consist of three individual projects: the Dark Energy Camera Legacy Survey (DECaLS), the Beijing-Arizona Sky Survey (BASS), and the Mayall z-band Legacy Survey (MzLS). These surveys utilized facilities such as the Blanco, Bok, and Mayall telescopes, supported by the National Science Foundation (NSF) and operated by different observatories including NSF’s NOIRLab.

We thank the respective teams and funding agencies for making these data publicly available. For detailed acknowledgments and funding information, please refer to the original publications and data release notes.
\end{acknowledgments}

\bibliography{desi_rings}{}
\bibliographystyle{aasjournalv7}



\end{document}